\documentclass[amsfonts, amssymb, amsmath, showkeys, nofootinbib, twocolumn, superscriptaddress, aps, letter]{revtex4-1}
\usepackage{amsthm}
\usepackage[english]{babel}
\usepackage[utf8]{inputenc}
\usepackage[colorinlistoftodos, color=green!40, prependcaption]{todonotes}
\usepackage{mathtools}
\usepackage{physics}
\usepackage{xcolor}
\usepackage{graphicx}
\usepackage{adjustbox}
\usepackage{placeins}
\usepackage[T1]{fontenc}
\usepackage{lipsum}
\usepackage{csquotes}
\usepackage{tikz}
\usetikzlibrary{quantikz2}
\usepackage{bm}% bold math
\usepackage{braket}
\usepackage{multirow}
\usepackage{siunitx}

\usepackage{hyperref}
\hypersetup{
    colorlinks=true,
    linkcolor=blue,
    urlcolor=blue,
    citecolor=blue
    }
\usepackage[capitalise]{cleveref}

\newcommand{\expect}[1]{\ensuremath{\left\langle#1\right\rangle}}
\newcommand{\ab}{\ensuremath{{}_{(a,b)}}}
\DeclareMathOperator{\arctantwo}{arctan2}
\DeclarePairedDelimiter\floor{\lfloor}{\rfloor}

\usepackage{color}
\definecolor{blue-violet}{rgb}{0.54, 0.17, 0.89}

\newcommand{\kcy}[1]{}

\begin{document}

\title{Heisenberg-limited calibration of entangling gates with robust phase estimation}

\author{Kenneth Rudinger}
    \thanks{These authors contributed equally to this work. Corresponding author: kmrudin@sandia.gov}
    \affiliation{Quantum Performance Laboratory, Sandia National Laboratories, Albuquerque, NM 87185 and Livermore, CA 94550}
\author{J. P. Marceaux}
    \thanks{These authors contributed equally to this work. Corresponding author: kmrudin@sandia.gov}
    \affiliation{Quantum Performance Laboratory, Sandia National Laboratories, Albuquerque, NM 87185 and Livermore, CA 94550}
    \affiliation{Graduate Group in Applied Science and Technology, University of California at Berkeley, Berkeley, CA 94720, USA}
\author{Akel Hashim}
    \affiliation{Quantum Nanoelectronics Laboratory, Department of Physics, University of California at Berkeley, Berkeley, CA 94720, USA}
\author{David I. Santiago}
    \affiliation{Computational Research Division, Lawrence Berkeley National Lab, Berkeley, CA 94720, USA}
\author{Irfan Siddiqi}
    \affiliation{Quantum Nanoelectronics Laboratory, Department of Physics, University of California at Berkeley, Berkeley, CA 94720, USA}
    \affiliation{Computational Research Division, Lawrence Berkeley National Lab, Berkeley, CA 94720, USA}
    \affiliation{Materials Sciences Division, Lawrence Berkeley National Lab, Berkeley, CA 94720, USA}
\author{Kevin C. Young}
    \affiliation{Quantum Performance Laboratory, Sandia National Laboratories, Albuquerque, NM 87185 and Livermore, CA 94550}

\date{\today} % Leave empty to omit a date

\begin{abstract}
    \noindent The calibration of high-quality two-qubit entangling gates is an essential component in engineering large-scale, fault-tolerant quantum computers. However, many standard calibration techniques are based on randomized circuits that are only quadratically sensitive to calibration errors. As a result, these approaches are inefficient, requiring many experimental shots to achieve acceptable performance. In this work, we demonstrate that robust phase estimation can enable high-precision, Heisenberg-limited estimates of coherent errors in multi-qubit gates. Equipped with an efficient estimator, the calibration problem may be reduced to a simple optimization loop that minimizes the estimated coherent error. We experimentally demonstrate our calibration protocols by improving the operation of a two-qubit controlled-Z gate on a superconducting processor, and we validate the improved performance with gate set tomography. Our methods are applicable to gates in other quantum hardware platforms such as ion traps and neutral atoms, and on other multi-qubit gates, such as CNOT or iSWAP. 
\end{abstract}


\maketitle

\section{Introduction} \label{sec:intro}
Errors in quantum logic operations are a leading barrier to practical, scalable quantum computation. Some errors, such as those arising from high-frequency noise in the environment or spontaneous emission from the physical qubit, are unavoidable without significant architectural changes to the system. Others, however, are simply \textit{calibration errors}---the result of tunable control parameters that are set suboptimally. In this work we demonstrate high-precision calibration of a two-qubit phase gate using robust phase estimation (RPE) \cite{Kimmel_2015}, an inexpensive characterization protocol that provides reliable estimates of the coherent gate errors that arise naturally from miscalibration. 

In gate-model quantum computers, logic operations (or, \textit{quantum gates}) are implemented by modulating the qubit's environment with drive waveforms, usually electromagnetic pulses, whose details and influence can generally be described with a small set of interpretable parameters. Some of these parameters are controllable, such as the desired timing, frequency, phase, or amplitude of the pulse. However, other parameters are neither controllable nor directly observable, such as instantaneous magnetic field strengths, temperature of acousto-optic modulators, or resonant frequencies of nearby two-level systems. Furthermore, these parameters are likely to fluctuate due to external factors, such as temperature variations or instrument drift. Frequent tuning of the controllable parameters is necessary to compensate for these changes in the qubit's environment or in the control hardware.

Most approaches to quantum gate calibration rely on iteratively adjusting controllable parameters until one or more proxies of gate performance are as close as possible to their optimal values. However, this is complicated by the fact that common performance proxies are not directly observable---ie., we cannot place a sensor to directly measure the local magnetic field at a qubit, nor can we directly observe a gate's fidelity or process matrix. Instead calibration information must be inferred by running some set of probe quantum circuits on a quantum device and (usually) post-processing the results. Exactly what proxies are selected and what protocols are used to measure them can significantly impact the time required for calibration and the quality of the resulting gates.

Small miscalibration errors largely manifest as \textit{coherent} errors that affect the unitary action of a quantum gates. Coherent errors are systematic and correlated in time, and their effects can be significantly amplified in deep, periodic circuits \cite{nielsen2021a}. But metrics based on \textit{infidelity}, perhaps the most common proxies for gate performance, are only quadratically sensitive to coherent errors (see supplementary material of \cite{madzik2022a}).
Despite this significant drawback, calibration routines based on fidelity proxies remain common.  For example, one relatively straightforward way to estimate gate fidelities is randomized benchmarking (RB) \cite{PRXQuantum.3.020357}; RB has therefore been used for calibrating quantum gates \cite{kelly2014optimal}.  Another example is the calibration of two-qubit gates via \textit{state} fidelity experiments \cite{PhysRevLett.127.200502, sheldon2016procedure, Sundaresan2020-sj}, including the ``parity scan'' popular for characterizing and calibrating two-qubit gates in trapped-ion quantum computers \cite{Monz2011-sj}.

Alternative approaches attempt to quantify coherent errors specifically. For example, process tomography \cite{Mohseni_2008} experiments can be used to directly measure the types and magnitudes of all of the errors that impact a gate, both coherent and incoherent. A calibration protocol based on process tomography, however, suffers both imprecise estimates and an abundance of nuisance parameters. Each quantum circuit in process tomography uses the target gate once, so the total shot count approximates the total number of gate uses in the experiment, $N$. The uncertainty of the process tomography estimator is bound by the standard quantum limit \cite{doi:10.1126/science.1104149} to decay as $\mathcal{O}(1/\sqrt{N})$. Long-sequence gate set tomography (GST) \cite{nielsen2021a} overcomes this by running deep circuits consisting of many repeated gate applications to amplify errors and achieve Heisenberg scaling; a GST experiment with maximum circuit depth $\mathcal{O}(L)$ can yield an estimate whose uncertainty scales as $\mathcal{O}(1/L)$. However, GST can require hundreds to thousands of unique circuits to achieve high-precision estimates of the target coherent errors. The large overhead of GST is due to the fact that it estimates a full description of the gate set -- including numerous incoherent parameters that have no bearing on calibration. Estimating all the parameters is costly and unnecessary.

In this work, we demonstrate calibration based robust phase estimation (RPE) that overcomes these issues. RPE was introduced in a single-qubit context to provide high-precision estimates of coherent errors, such as rotation and axis errors in single-qubit gates \cite{Kimmel_2015,Rudinger2017a}. Its eponymous robustness is achieved by its particular data analysis procedure, which is insensitive to errors in state preparation and measurement, as well as incoherent (stochastic) errors. Its long sequences further allow for Heisenberg-limited uncertainty in estimates of coherent errors. As we demonstrate below, RPE extends straightforwardly to multi-qubit gates and can be used as the basis of a high-precision calibration protocol. 

\begin{figure*}[ht]
    \centering
    \includegraphics[width=\textwidth]{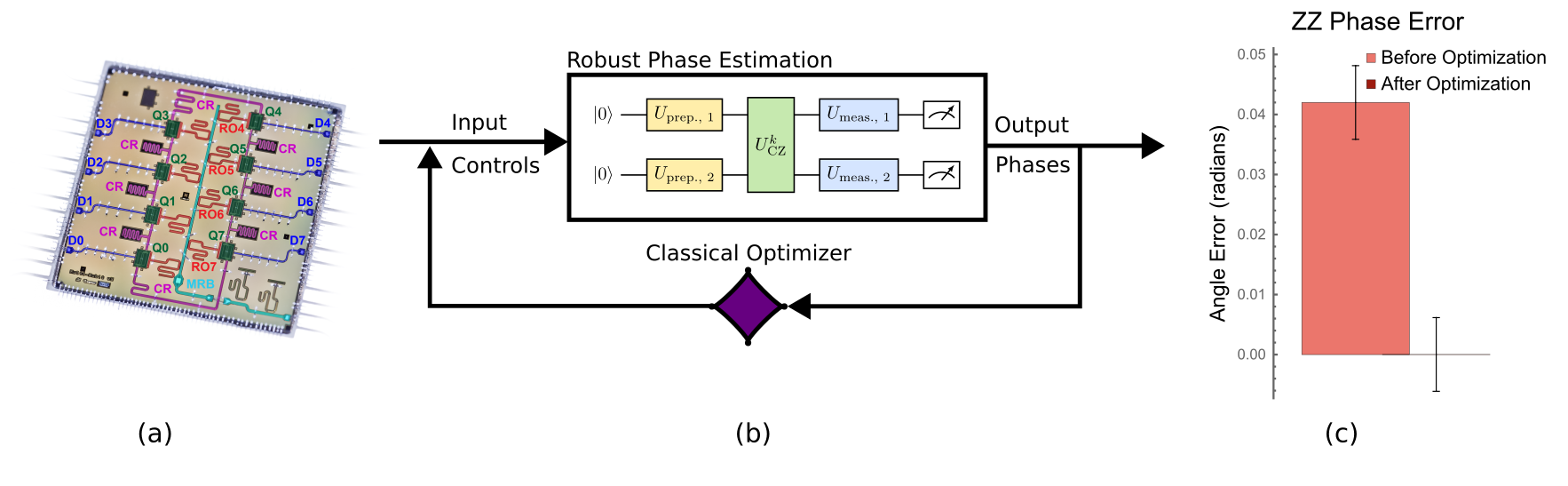}
    \caption{Components of RPE-based calibration. Panel (a) shows the layout of the superconducting chip used in our experiment. Eight qubits (green) are arranged in a ring geometry and are coupled to nearest neighbors via fixed-frequency coupling resonators (CR, purple). Each qubit has an independent drive line (blue) and are coupled to independent readout resonators (RO, red), which can be measured simultaneously via a multi-plexed readout bus (MRB, cyan). Panel (b) depicts RPE calibration as an input-output map between controsl and estimated phases. A classical optimizer uses estimated phases from particular control inputs to identify the optimal controls that minimize the phase error. Panel (c) depicts the $ZZ$ phase error on the controlled-Z gate before and after optimization. We see that a classical optimizer is able to identify the optimal controls that minimize the $ZZ$ phase error on the controlled-Z gate.}
    \label{fig:main_fig}
\end{figure*}

Equipped with multi-qubit RPE as an efficient and high-quality estimator of coherent error, the calibration problem reduces to a problem of classical control theory. The quantum system can be modeled as an input-output map between control inputs (experimental knobs like microwave power) and estimated outputs (the phases estimated by RPE); see Figure \ref{fig:main_fig}(b). Numerous classical methods are available to maximize gate performance in this setting, and we employ a simple classical optimizer that minimizes a cost function between the ideal and estimated phases. More sophisticated optimization approaches are possible, and the best choice will likely be specific to a particular experimental platform, control hardware, and target gate.

We apply our RPE-based calibration to a controlled-Z (CZ) gate on the superconducting quantum processor shown in Fig.~\ref{fig:main_fig}(a). The CZ entangling gate is widely used both in near-term, intermediate-scale quantum (NISQ) algorithms \cite{Mi2022-rj} and circuits employing quantum error correction \cite{Martinis2015-ps}. Indeed, achieving high two-qubit gate fidelity is one of the most important milestones to a large-scale, fault-tolerant quantum computer. Our methods proves to be a simple and inexpensive procedure to optimize native CZ gate control parameters on current hardware, with results depicted in Fig.~\ref{fig:main_fig}(c). While our experiment was run specifically on a superconducting transmon system, our methods are hardware-agnostic and may be straightforwardly deployed on other gate-based quantum hardware. Furthermore, simple modifications of our approach also permit the characterization and calibration of other entangling gates, such as controlled-NOT (CNOT). 

This manuscript is organized as follows. In Sec.~\ref{sec:rpe} we review single-qubit RPE and develop the theory of multi-qubit RPE with application to the controlled-Z gate in transmon systems. Sec.~\ref{sec:experiment} discusses experimental implementation of RPE-based calibration of a CZ gate between two transmon qubits. Here we analyze the performance improvement by comparing gate performance metrics collected before and after calibration. Sec.~\ref{sec:conclusions} concludes with a discussion and outlook. 

\section{Robust Phase Estimation} \label{sec:rpe}

\noindent Robust phase estimation (RPE) was first introduced in \cite{Kimmel_2015} as a enhancement of the Heisenberg-limited phase estimation protocol of \cite{higgins2009a} that added robustness both to incoherent noise in the logical gates and to errors in state preparation and measurement. Its original formulation estimated the unique relative phase of a single-qubit gate, but has since been expanded to measure relative phases between arbitrary eigenstates in a two-qubit gate \cite{Russo2021a}. As discussed in Appendix \ref{apendix:Heisenberg}, the multi-qubit generalization of RPE inherits the Heisenberg-limited scaling of single-qubit RPE.

Algorithms for performing phase estimation can generally be divided into two classes: entanglement-assisted and entanglement-free. Entanglement-assisted protocols make use of ancilla qubits and are often used as primitives in large quantum algorithms, especially in quantum chemistry \cite{Martyn_2021}. Entanglement-free protocols do not require additional ancilla systems and are often used in metrological or tomographic applications. These algorithms achieve high-precision by repeated application of the unitary operator. RPE, as discussed in this manuscript, is an entanglement-free protocol, which is most appropriate for tuning noisy quantum computers that may be (initially) moderately far from the ideal operations.

To introduce the theory of RPE, it is most straightforward to begin with the spectral decomposition of a unitary matrix $\mathbf{U}$ on a $d$-dimensional state space:
\begin{equation}
    \mathbf{U} = \sum_{a} e^{i \phi_a} \ketbra{{E_a}}, 
\end{equation}
where $\{e^{i \phi_a}\}_{a=1}^{d}$ are the eigenvalues of $\mathbf{U}$ and $\{\ket{{E_a}}\}_a$ are the associated eigenstates. We refer to $\phi_k$ as an \textit{eigenphase} of $\mathbf{U}$. Because global phases are unobservable, the eigenphases themselves are not uniquely defined and cannot be individually resolved. However, differences between two eigenphases are physically observable, and we refer to $\phi\ab = \phi_a - \phi_b$ as a \textit{relative phase}. The goal of phase estimation is to estimate one or more of these relative phases.

To estimate the relative phase $\phi\ab$, RPE makes use of the following four states:
\begin{align}
    \ket{\pm x\ab} = \frac{1}{\sqrt{2}} \left( \ket{{E_a}} \pm \ket{{E_b}} \right) \\
    \ket{\pm y\ab} = \frac{1}{\sqrt{2}} \left( \ket{{E_a}} \pm i \ket{{E_b}} \right)
\end{align}
From these states we can define two operators, an in-phase operator $I\ab$ and a quadrature operator $Q\ab$:
\begin{align}
    I\ab &= \ketbra{x\ab} - \ketbra{-x\ab}\\
    Q\ab &= \ketbra{y\ab} - \ketbra{-y\ab}
\end{align}
The RPE protocol begins by creating the initial superposition state 
\begin{equation}
    \ket{+x\ab} = \frac{1}{\sqrt{2}} \left( \ket{E_a} + \ket{E_b}\right).
\end{equation}
A depth parameter $k$ is initialized to 0, and the initialized state is subjected to $2^k$ applications of the unitary $\mathbf{U}$, resulting in
\begin{equation}
    \mathbf{U}^{2^k} \ket{+x\ab} = \frac{1}{\sqrt{2}} \left( e^{i 2^k\phi_a}\ket{E_a} + e^{i 2^k\phi_b}\ket{E_b}\right).
\end{equation}
The state is then measured in the appropriate bases to infer the expectation values of the in-phase and quadrature operators, $\expect{I\ab(2^k)}$ and $\expect{Q\ab(2^k)}$, as functions of the number of applications of the unitary. (As $I_{(a,b)}$ and $Q_{(a,b)}$ do not commute, each value of $k$ yields two distinct quantum circuits to run.)  
This process is repeated for $k\in[0,1,\ldots,k_{\max}]$.  If no real-time feedback is used, $k_{\max}$ is determined in advance.  However, if real-time feedback is deployed, then $k$ continues to increment by 1 until a consistency check (see below) fails.
In the absence of measurement error or decoherence, $\expect{I\ab(2^k)} = \cos(2^k \phi_{a,b})$ and $\expect{Q\ab(2^k)} = \sin(2^k\phi_{a,b})$. However, when $d>2$, errors in state preparation, measurement, or in the gate action itself can push state population outside the $\left\{\ket{E_a}, \ket{E_b}\right\}$ subspace. Observation of such ``leakage'' states will not inform the phase estimate. We therefore post-select the measurement outcomes when constructing an empirical estimate of the expectation values. For instance, when measuring the in-phase operator, we define $N_{\pm x}$ as the number of times $\ket{\pm x\ab}$ is observed. The post-selected expectation value is simply:
\begin{equation}
    \expect{I\ab(2^k)}_p = \frac{N_{+x} - N_{-x}}{N_{+x} + N_{-x}}.
\end{equation}

RPE produces a sequence of increasingly precise estimates of the relative phase as:
\begin{equation}\label{eq:rel_phase}
    \hat\phi_{a,b}^{(k)} 
    = \frac{1}{2^k}\left(\arctantwo\left( \expect{Q\ab(2^k)}_p,\expect{I\ab(2^k)}_p \right) 
        + 2\pi n_k\right),
\end{equation}
where $\arctantwo$ is the 2-argument arctangent function \cite{Russo2021a}, defined so that $\arctantwo(\Im(e^{i\phi}), \Re(e^{i\phi})) = \phi$ when $\phi \in (0,2\pi]$, and  $n_k$ is an integer that selects the proper branch of the arctangent function. We assume $n_0$ can be chosen unambiguously. For most standard quantum gates, $n_0=0$. For $k>0$, $n_k$ is chosen to minimize the angular difference between the estimate $\hat\phi_{a,b}^{(k)}$ and its immediate predecessor $\hat\phi_{a,b}^{(k-1)}$.  An explicit form for $n_k$ is given in Appendix \ref{apn:n_k}. 

The eponymous \textit{robustness} of RPE promises that as long as the effects out-of-model error do not exceed a threshold, then the estimates produced from the RPE procedure will be within an error bound that decreases exponentially as the number of applications of the unitary. Here, out-of-model errors include any error process that does not simply appear as changes in the relative phase, including state preparation and measurement (SPAM) error, incoherent effects in the gate, and shot noise uncertainty. Formally, the promise is that as long as
\begin{align}
\frac{1}{2}\abs{\expect{I\ab(2^k)}_p - \cos(2^k \phi_{a,b})} &< \sqrt{\frac{3}{32}},\\
\frac{1}{2}\abs{\expect{Q\ab(2^k)}_p - \sin(2^k \phi_{a,b})} &< \sqrt{\frac{3}{32}}
\end{align}
then the root-mean-squared error for the estimated phase $\hat{\phi}^{k}\ab$ at depth $2^k$ is no more that $\pi/2^{k+1}$ \cite{Russo2021b,Rudinger2017a}.

While of great theoretical utility, the robustness bounds measure the distance between estimates and ideal values that cannot be known experimentally. Incoherent effects in a gate will build up with the number of applications of the unitary and at some point the robustness bounds will be violated. To turn RPE into an experimentally robust protocol, a suite of consistency checks have been developed \cite{Russo2021b}. We employ the ``angular historical'' consistency check here that defines an operational uncertainty range (\textit{not} a statistical uncertainty range) surrounding the estimate:
\begin{equation}
    \Delta_\phi^{(k)} = \left[ \hat \phi_k - \frac{1}{2^{k+1}}\frac{\pi}{3}, \hat\phi_k + \frac{1}{2^{k+1}}\frac{\pi}{3}\right]
\end{equation}
The consistency check used to accept generation $k$ ensures that the generation $k$ estimate lies in the intersection of all previous uncertainty ranges: 
\begin{equation}
    \hat\phi_{k} \in \bigcap_{k^\prime < k} \Delta_\phi^{(k^\prime)}
\end{equation}
If the estimate does not lie in this set, then the analysis is terminated at generation $k-1$.

Because multi-qubit RPE involves more than one non-trivial phase, we require multiple experiments to fully characterize the spectrum of a multi-qubit unitary.  This requires performing a set of RPE experiments that both measures a linearly independent set of eigenphase differences and is also complete (each eigenphase appears in at least one eigenphase difference).  Therefore complete phase estimation of a $d$-dimensional unitary requires $d-1$ RPE experiments.  How to choose which $d-1$ eigenphase differences to measure (out of a possible $\binom{d}{2}$ choices) depends on which pairs $(E_a,E_b)$ yield easy-to-prepare $\ket{+x_{(a,b)}}$ states and easy-to-implement $I_{(a,b)}$ and $Q_{(a,b)}$ measurements.  This will generally be a function of the gate in question. 

\subsubsection{Controlled-Z RPE}
\label{subsec:controlled-Z-rpe}

We now specialize our discussion to the specific controlled-Z (CZ) RPE experiment design employed in this work. The CZ gate has a target unitary operation $\mathbf{U}_{CZ} = \mathbf{I} \otimes \ketbra{0}{0} + \mathbf{Z} \otimes \ketbra{1}{1}$, or written in an equivalent exponential notation
\begin{equation}
    \mathbf{U}_{CZ} = \exp \bigg( -\frac{i}{2} \bigg(-\frac{\pi}{2} \mathbf{II} + \frac{\pi}{2} \mathbf{ZI} + \frac{\pi}{2} \mathbf{IZ} - \frac{\pi}{2} \mathbf{ZZ} \bigg) \bigg).
\end{equation}

An experimental implementation of the CZ unitary operation will be a noisy approximation of the target (ideal) operation. Coherent errors in the unitary operator's implementation will change the eigenvalues of the operator and hence are measurable with RPE. Our approach to calibrate the CZ gate is to assume a simple model with free parameters for the Pauli coefficients of $\mathbf{IZ}, \mathbf{ZI}$, and $\mathbf{ZZ}$. This model is physically motivated by the off-resonant nature of the drive (see Section \ref{sec:experiment}) but only approximately captures all coherent errors (see Fig.~\ref{fig:angle_errors}(a)). We use RPE to estimate the coefficients of each of these Pauli operators and optimize over control inputs to set the estimated parameters as close to their target value as possible. The specific model we consider is 
\begin{equation}\label{eq:cz_unitary}
    \tilde{\mathbf{U}}_{CZ} = \exp \bigg(- \frac{i}{2} (  \theta_{ZI} \mathbf{ZI} + \theta_{IZ} \mathbf{IZ} + \theta_{ZZ} \mathbf{ZZ} ) \bigg), 
\end{equation}
where we ignore the physically unobservable coefficient on $\mathbf{II}$. $\theta_{IZ}$ and $\theta_{ZI}$ have a target value of $\pi/2$, and $\theta_{ZZ}$ has a target value of $-\pi/2$ 

We start by identifying the eigenstates and eigenphases of the target unitary for the CZ gate, listed in Table~\ref{tab:eigensys}. To measure the eigenphases, we construct three classes of RPE experiments that consist of a particular state preparation, application of the CZ gate according to a logarithmically-spaced interval, and finally a particular measurement in both the $I$ and $Q$ bases. Each individual RPE experiment measures a particular linear combination of the model parameters, and we list the details of each experiment in Table \ref{tab:three_experiments}. After running RPE analysis on the experimental data collected from the three experiments outlined in the previous paragraph, we learn a set of three relative phases $\phi_{00, 01}, \phi_{10, 11}, \phi_{01, 11}$ that are related to the model parameters according to the following linear relationship: 

\begin{table}[t]
    \centering
    \begin{tabular}{|c |c |}
        \hline
         Eigenstate & Eigenphase \\
         \hline
         $\ket{00}$ & $\theta_{ZI} + \theta_{IZ} + \theta_{ZZ}$  \\
         $\ket{01}$ & $\theta_{ZI} - \theta_{IZ} - \theta_{ZZ}$  \\
         $\ket{10}$ & $-\theta_{ZI} + \theta_{IZ} - \theta_{ZZ}$ \\
         $\ket{11}$ & $-\theta_{ZI} - \theta_{IZ} + \theta_{ZZ}$ \\
         \hline
    \end{tabular}
    \caption{ Eigenstates and eigenphases for the model CZ operation, Eq.~\ref{eq:cz_unitary}. The $II$ phase was set to 0 using $U(1)$ phase freedom.}
    \label{tab:eigensys}
\end{table}

\begin{equation}\label{eq:linear_system}
    \begin{bmatrix}
        \phi_{00, 01} \\ 
        \phi_{10, 11} \\
        \phi_{01, 11} \\
    \end{bmatrix} 
    =
    \begin{bmatrix}
        1 & 0 & 1 \\
        1 & 0 & -1\\
        0 & 1 & -1\\
    \end{bmatrix}
    \begin{bmatrix}
        \theta_{IZ} \\ 
        \theta_{ZI} \\
        \theta_{ZZ}
    \end{bmatrix}. 
\end{equation}
We invert the linear system to find the actual estimate of the model parameters.

\begin{table*}[]
    \centering
    \renewcommand{\arraystretch}{2}
    \begin{tabular}{|c|c | c | c|}
    \hline
    State prep. & $I\ab$ & $Q\ab$ & Relative phase \\
    \hline
    $\ket{0+}$ & $\ketbra{0+}{0+}-\ketbra{0-}{0-}$ & $\ketbra{0+_y}{0+_y}-\ketbra{0-_y}{0-_y}$ &  $\phi_{00, 01} = \theta_{IZ} + \theta_{ZZ}$ \\
    $\ket{1+}$ & $\ketbra{1+}{1+}-\ketbra{1-}{1-}$ & $\ketbra{1+_y}{1+_y}-\ketbra{1-_y}{1-_y}$ &  $\phi_{10, 11} = \theta_{IZ} - \theta_{ZZ}$ \\
    $\ket{+1}$ & $\ketbra{+1}{+1}-\ketbra{-1}{-1}$ & $\ketbra{+_y1}{+_y1}-\ketbra{-_y1}{-_y1}$ &  $\phi_{01, 11} = \theta_{ZI} - \theta_{ZZ}$ \\
    \hline
    \end{tabular}
    \caption{The state preparations and $I\ab$ and $Q\ab$ observables that are used to estimate the relative phases that constitute a CZ gate.  Each RPE circuit is formed by preparing in a state given in the ``State prep.'' column, followed by an application of $2^k$ CZ gates, followed by measuring either the $I\ab$ or $Q\ab$ observable.  For explicit circuit diagrams, see Tab. \ref{tab:circuit_table}.}
    \label{tab:three_experiments}
\end{table*}

In summary, the RPE design we use to calibrate the CZ gate in this work consists of three independent RPE experiments. We collect empirical distributions for our three experiments and renormalize with post-selection. The three experiments measure three phases that are linear functions of the parameters of a calibration model, and we invert the linear system to find estimates of the model parameters. We optimize over control inputs to set the estimated parameters as close as possible to their target values, using the last trusted generation as our final estimate.

In our analysis, we employ a \textit{commutative model} that assumes the unitary gates are generated by a commuting set of Pauli operators (\textit{i.e.}, all Pauli terms in the gate's Hamiltonian commute). In the case of commutative Hamiltonian terms, the eigenvectors of the resulting unitary representation are independent of particular parameter values, which greatly simplifies RPE analysis. It should be possible to extend our techniques to the non-commuting case in a limit of small errors, similarly to the single-qubit axis angle estimator constructed in \cite{Kimmel_2015}. However, the multi-qubit Pauli algebra exhibits complicated structure, which means that analytic derivation of the estimators for such sequences will be difficult. Future work will address this problem. It is often the case that the commutative approximation will capture the dominant sources of error, as in this work.

\section{Experiment}\label{sec:experiment}

We deploy RPE to calibrate the phases of a CZ gate. Our hardware consists of an 8-qubit transmon quantum processor with fixed-frequency qubits and fixed-frequency coupling resonators at the the Advanced Quantum Testbed \cite{PhysRevLett.127.200502}. Our CZ gate is realized using off-resonant drives on both qubits to induce a conditional AC Stark shift \cite{PhysRevLett.127.200502, wei2021quantum}. As in Eq.~\ref{eq:cz_unitary}, the native Hamiltonian for this interaction consists of an entangling $\mathbf{ZZ}$ term, as well as local $\mathbf{IZ}$ and $\mathbf{ZI}$ phases that are implemented virtually \cite{McKay_2017} following the gate pulse. As reported in \cite{PhysRevLett.127.200502}, this CZ gate can operate over a wide range of parameters. This tunability is beneficial on fixed-frequency, fixed-coupling processors \cite{nguyen2022programmable}, but it also makes finding the optimal parameter regime difficult, as one must calibrate over potentially many different parameters. While the calibration procedure introduced in \cite{PhysRevLett.127.200502} has been used to calibrate high-fidelity gates as measured via randomized benchmarks, a fine-grained characterization of the angle errors in the gate was not performed in \cite{PhysRevLett.127.200502}.

Due to the off-resonant nature of the CZ drive, the three Hamiltonian terms $\mathbf{ZZ}$, $\mathbf{IZ}$, and $\mathbf{ZI}$ describe the dominant effect of the drive (see Eq.~\ref{eq:cz_unitary}). We use RPE to optimize and calibrate all three terms.  Other interactions such as $\mathbf{YI}$ or $\mathbf{ZX}$ coupling may also develop during the pulse via a residual resonant or cross-resonant interactions \cite{sheldon2016procedure}, but their contributions to the total error are expected to be small compared to $\mathbf{ZZ}$, $\mathbf{IZ}$, and $\mathbf{ZI}$. These terms do not affect the eigenphases to first order, so they are not measured by RPE. GST results (see Fig.~\ref{fig:angle_errors}(a)) verify this expectation. To calibrate the $\theta_{ZZ}$ phase, we use a classical optimizer to find the optimal amplitude and frequency of the drive pulse which minimizes a cost function between the estimated and target $\theta_{ZZ}$ phases. The local $\theta_{ZI}$ and $\theta_{IZ}$ phases are straightforward to measure via RPE and correct with virtual $\mathbf{Z}$ rotation gates. We report measurable improvement in the gate's angle errors, fidelity, and diamond-distance to the target operation based on gate set tomography experiments.  We use the open-source software packages \texttt{pyRPE} \cite{pyRPE} and \texttt{pygsti} \cite{pygsti} to perform the RPE and gate set tomography calculations, respectively.

\subsection{Coarse grained calibration}

An initial set of operational parameters can be identified through a coarse-grained calibration routine for the CZ gate \cite{PhysRevLett.127.200502}. There are two steps in the coarse-grained calibration: (1) determine the amplitude and frequency settings for the drive pulses and (2) determine the local phases on the drive pulses. These steps, summarized below, result in a decent approximation of the target gate but that still has significant coherence error that can be further reduced with RPE calibration.

(1) \textit{$\theta_{ZZ}$ Calibration:} The optimal frequency of the CZ gate typically resides between the GE transition of the higher frequency qubit and the EF transition of the lower frequency qubit, termed the \textit{straddling regime}. A coarse 2D sweep in amplitude and frequency within the straddling regime can be used to find the operating parameter region of the gate. The optimal amplitude and frequency parameters are determined by finding the region with the highest \textit{conditionality},
\begin{equation}\label{eq:cond}
    R = \frac{1}{2} || \textbf{r}_0 - \textbf{r}_1 ||^2 ,
\end{equation}
where $\textbf{r}_0$ and $\textbf{r}_1$ are tomographic reconstructions of the target qubit's Bloch vector when prepared in $\ket{+}$ conditioned on preparing the control qubit in $\ket{0}$ or $\ket{1}$. Note that if one assumes that the effective Hamiltonian contain only $\mathbf{ZZ}$, $\mathbf{IZ}$, and $\mathbf{ZI}$ interactions, then one need only measure the $X$ and $Y$ components of each $r_i$ as the $Z$ components will not change as a function of the control setting. Here, $R$ is the Euclidean distance between the two conditional states implemented by the CZ gate. Regions of high conditionality correspond to regions of high entanglement, thus the amplitude and frequency that correspond to a large $R$ value directly relate to the $\mathbf{ZZ}$ term in the Hamiltonian.

(2) \textit{$\theta_{IZ}$ and $\theta_{ZI}$ Calibration:} A final, straightforward calibration (detailed in Appendix X of \cite{PhysRevLett.127.200502}) sets the local phases of the two drives that control the $\theta_{IZ}$ and $\theta_{ZI}$ phases of the gate. To summarize, let us say that the two qubits involved in the gate are A and B. To calibrate the local phase on the drive on A that controls the $\theta_{ZI}$ unitary phase, one prepares A in $\ket{+}$ and B in $\{\ket{0}, \ket{1}\}$. One measures the transition probabilities $|\bra{+ b} U_\text{CZ} \ket{+ b}|^2$ for $b \in \{0,1\}$ as a function of the local drive phase on A, and sets the phase such that for $b=0$ the response is maximal and $b=1$ the response is minimal.

\subsection{RPE Calibration}

\begin{figure*}[ht]
    \centering
    \includegraphics[width=\textwidth]{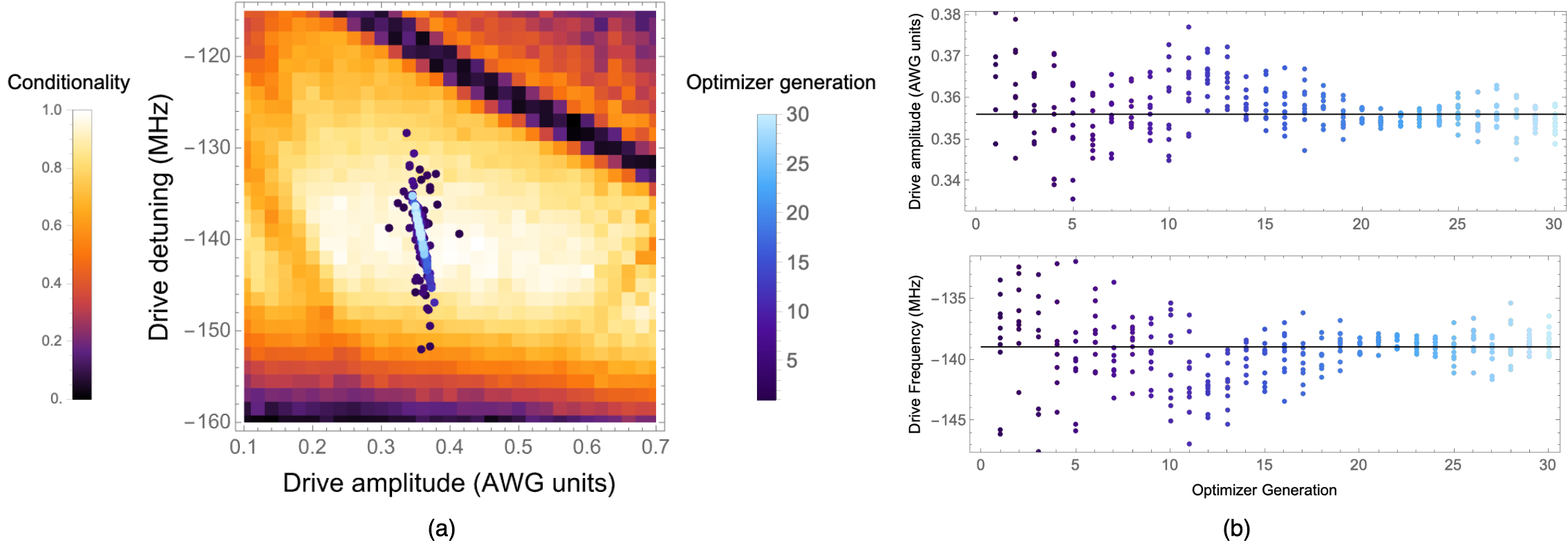}
    \caption{Cost function landscape exploration and optimizer trajectories. Panel (a) depicts a coarse-grained sweep of the conditionality landscape (Eq. \ref{eq:cond}) and the exploration of the RPE cost function (Eq. \ref{equ:cost}) as points embedded in the conditionality landscape. Regions of high conditionality correspond to near-optimal gate performance as discussed in the text surrounding Eq.~\ref{eq:cond}, but the conditionality is a low-precision metric that is not Heisenberg scaling. The RPE cost function provides a significantly more precise estimate of the performance. The initial conditionality sweep serves to identify a reasonable range of control parameters for high-precision RPE optimization. Panel (b) depicts the control trajectory chosen by the optimizer to minimize the error in the $\theta_{ZZ}$ phase. The black central lines indicate the final control parameters chosen by the optimizer. One can observe from Panel (a) that the optimizer identifies a 1-dimensional subspace in the 2-dimensional control parameter space, and the majority of the exploration is done in that one dimensional, indicating that the optimizer is performing a type of principle component analysis and optimizing over only the most significant component. }
    \label{fig:landscape}
\end{figure*}

After performing a 2D sweep in frequency and amplitude for regions of high conditionality, we use RPE along with a classical optimizer to find the optimal operating parameters. The optimizer searches the amplitude-frequency landscape for the parameters that minimize the angle error in $\theta_{ZZ}$. The angles errors in $\theta_{IZ}$ and $\theta_{ZI}$ are finally corrected with virtual Z gates after finding an optimal amplitude-frequency pair. The experimental designs provided in Section \ref{subsec:controlled-Z-rpe} and summarized in Table \ref{tab:three_experiments} provide the high-precision estimates of the $\theta_{ZZ}$, $\theta_{IZ}$, and $\theta_{ZI}$ phases. Our RPE technique should be compared to fine-tuning these parameters using pulse-amplified CZ sequences and maximizing the conditionality. However, calibrating based on 2D sweeps is inefficient, requiring many more measurements than necessary, and furthermore, parameter sweeps scale very poorly as the number of interdependent parameters increases. 

We optimize drive parameters by exploring a cost function landscape over a bounded range of operating parameters, as shown in Fig.~\ref{fig:landscape}. We use a cost function that is the absolute difference between the target $\theta_{ZZ}$ phase of $-\tfrac{\pi}{2}$ and the RPE estimate $\hat{\theta}_{ZZ}$
\begin{equation}\label{equ:cost}
    J(\hat{\theta}_{ZZ}) = \abs{\theta_{ZZ} + \frac{\pi}{2}}. 
\end{equation}
We do not include the single qubit $\theta_{IZ}$ and $\theta_{ZI}$ phases in the cost function as these can be corrected with virtual Z gates after optimization. Our use of the absolute value, instead of the squared norm, is motivated by the fact that we wish to minimize the diamond distance to the target operation, and diamond distance is linearly sensitive to coherent errors. 

We employ the Covariance Matrix Adaptation Evolution Strategy (CMA-ES) \cite{hansen1996adapting, hansen1997convergence} for optimization. In each iteration loop, the optimizer selects 10 points at random in the landscape within the initial window. RPE provides estimates the $\theta_{IZ}$, $\theta_{ZI}$ and $\theta_{ZZ}$ phases for each of the 10 points, and a cost is assigned based on the $ZZ$ phase per Eq.~\ref{equ:cost}. From these 10 points, the optimizer narrows the search window, and the process is repeated until convergence or after a specific number of repetitions. See Fig.~\ref{fig:landscape}. After retrospectively analyzing the performance of the CMA-ES optimizer, we believe that there are better choices of optimization routines, and future work should be dedicated to identifying the best optimizer for RPE calibration. In particular, the CMA-ES optimizer works best in regimes with a large number of parameters, but in our case we had only two parameters to optimize.

After the optimizer explores the cost landscape, we obtain an optimal amplitude-frequency pair for the CZ drive pulse. The final step is to re-calibrate the local phases experienced by each of the qubits.  In order to ensure that each qubit experiences a local Z rotation of $\tfrac{\pi}{2}$ during the CZ, we use our knowledge of $\theta_{IZ}$ and $\theta_{ZI}$ to insert virtual local Z rotations of phases $\phi_{IZ}$ and $\phi_{ZI}$ on the two qubits, respectively.  These phases are chosen such that 
\begin{align}
    \phi_{IZ} = \frac{\pi}{2} - \theta_{IZ}\\
    \notag
    \phi_{ZI} = \frac{\pi}{2} - \theta_{ZI}
\end{align}
which fixes the errors in the local $Z$ phases.

\subsection{Results}

\begin{figure}
    \centering
    \includegraphics[width=\linewidth]{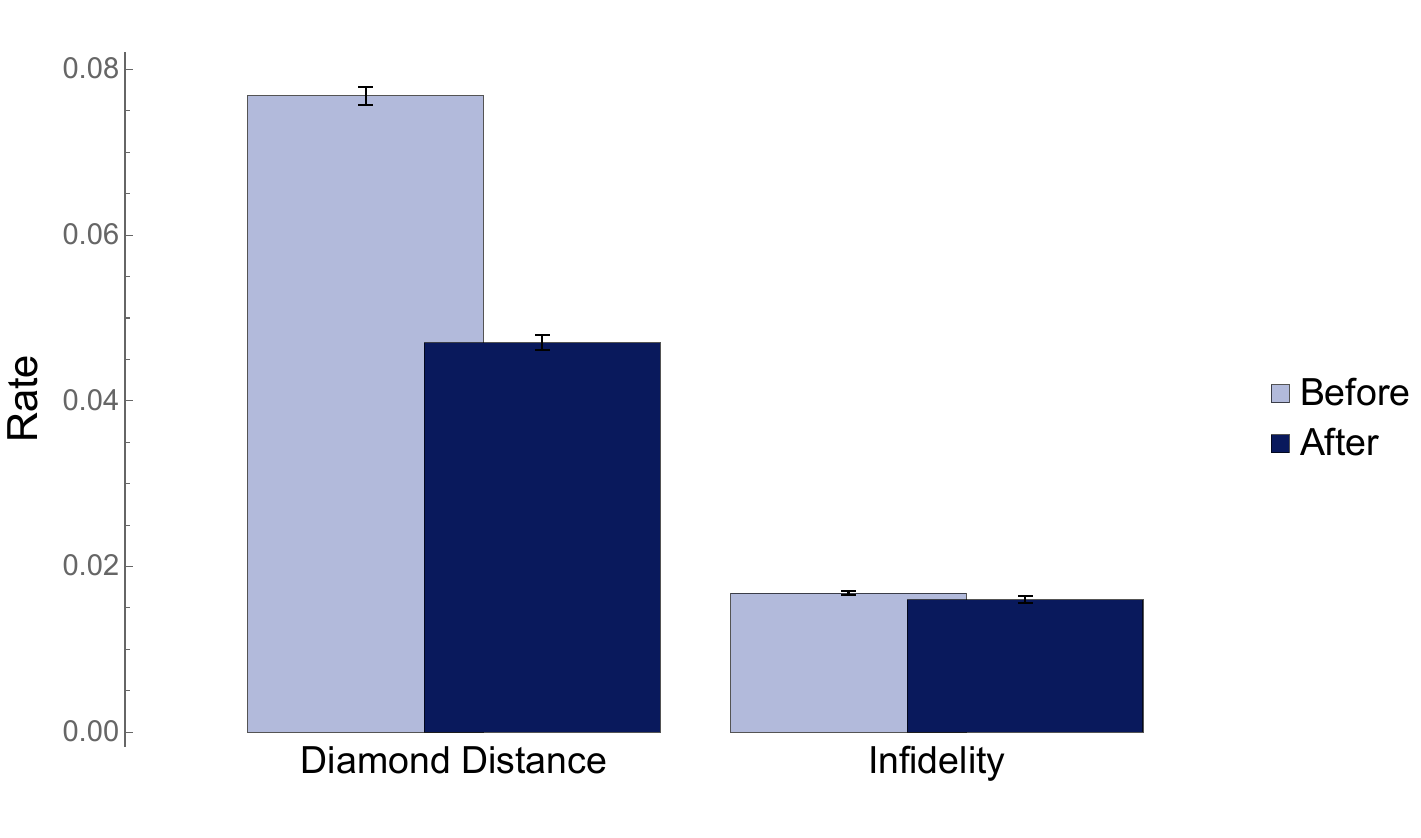}
    \caption{Diamond distance and infidelity before and after calibration. These metrics were measured with GST and their uncertainties determined by bootstrapping. Observe that calibration results in a significant reduction in diamond distance, but only a trivial reduction in infidelity. This fact highlights the ineffectiveness of randomized benchmarking numbers as a calibration metric, see the the main text for discussion.}
    \label{fig:metrics}
\end{figure}

To evaluate the performance of our calibration routine, we compare gate set metrics--measured via GST and shown in Fig.~\ref{fig:metrics}--that are based on experimental data collected before and after applying the RPE calibration. Our most significant improvement is in the angle error as measured by RPE, where we found that the calibrated gate improved by a factor of over 1000 (with caveat that the decoherence-limited uncertainty region is large). We found that the diamond distance of the CZ to its target operation improved by about a factor of $\sim2\times$. We found no significant reduction in infidelity, indicating persistent incoherent error.

\begin{figure*}[t]
    \centering
    \includegraphics[width=\textwidth]{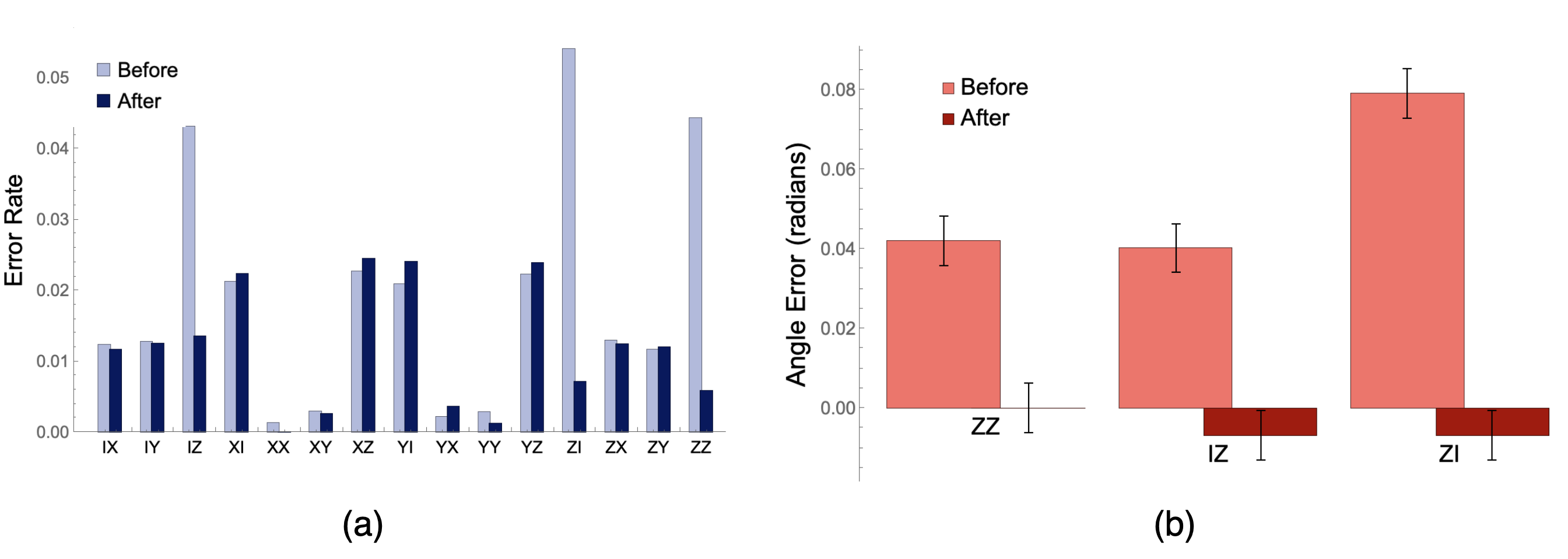}
    \caption{Metrics before and after calibration. Panel (a) shows the during-gate Hamiltonian error rates and panel (b) shows the angle error. The Hamiltonian error was measured with GST and indicate that significant resonant and cross-resonant terms persist in the drive Hamiltonian after calibration, which motivates future work to efficiently estimate and calibrate such small non-commuting terms. The angle error was measured with RPE, and the error bars correspond to the standard root-mean-square (RMS) error regions of an RPE estimate ($\pi/(2 d)$ for RPE of depth $d$). While RPE does estimate remaining errors in the local $\theta_{IZ}$ and $\theta_{ZI}$ phases after calibration, the magnitude of the errors falls within the RMS bounds of RPE.}
    \label{fig:angle_errors}
\end{figure*}

Ideally, we would expect the diamond distance to improve as much as the error in the $\theta_{ZZ}$ angle, since diamond distance is saturated by coherent errors in the limit of small stochastic gate noise \cite{hashim2023benchmarking}. However, we observe only a modest reduction in the diamond distance of the CZ gate. This is due to the fact that there are other error terms in the gate that are not measured by RPE which become important as we reduce the error in $\theta_{ZZ}$.

To measure the residual error remaining after calibration, we plot the during-gate Hamiltonian error generator rates in Fig.~\ref{fig:angle_errors}(a). The during-gate error generator is the logarithm of the measured process matrix minus the logarithm of the target \cite{blumekohout2022a}:
\begin{equation}
    L = \ln P - \ln \tilde{P}
\end{equation}
where $P$ is the measured process matrix and $\tilde{P}$ is the target process. In this way, $P = \exp(L + \tilde{L})$ where $\tilde{L}$ is the target error generator $\ln \tilde{P}$. We observe small coherent errors outside the $\{\theta_{ZI}, \theta_{IZ}, \theta_{ZZ}\}$ model that become significant after calibration. These terms were not addressed by the current calibration and contribute to the overall diamond norm error.

As we have said, the infidelity of the gate hardly improved despite substantial improvements in the coherent errors in the gate. We attribute this to the fact that the infidelity is only quadratically sensitive to coherent errors while linearly sensitive to stochastic errors \cite{blumekohout2022a,madzik2022a}. Hence, the gate's infidelity is dominated by stochastic errors. This fact also highlights why randomized benchmarking (RB) is not particularly well-suited for calibration and parameter optimization: the RB numbers measured in experiment are not very sensitive to improvements in the coherent error of a gate. Furthermore, an RB number is a single metric that depends on the distance between all the parameters of the experimental gate set and the target gate set, so one cannot determine if a reduction in RB numbers is due to improvement in the $\theta_{ZZ}$ phase or in local $\theta_{ZI}$ or $\theta_{IZ}$ phases.

In terms of timescales, the entire calibration experiment took approximately 3.75 hours in total. The time spent on classical optimization was only 8 minutes. An individual RPE run took about 45 seconds, so a single iteration in the optimization loop for all 10 point estimates took about 450 seconds. 30 optimization loops were made in total. About 50\% of the total time was spent compiling waveforms. We used a arbitrary waveform generator (AWG) to produce control pulses, and this default AWG has to recompile the entire waveform every time even a single parameter is changed. We expect that we could reduce the required calibration time by at least half by switching to FPGA-based control hardware, such as the \texttt{QubiC} system \cite{xu2021qubic, xu2023qubic, fruitwala2024distributed}, that can make rapid changes to parameters without the need to recompile an entire waveform. We expect that we can also reduce the time cost even further by employing better optimization strategies and reducing the number of shots per circuit to achieve faster convergence. In our case, no hyper-parameter tuning of the optimizer was performed. Future work should identify the best optimization strategies for RPE calibration.

\section{Conclusions and outlook} \label{sec:conclusions}

In this work, we detailed experiment design principles for multi-qubit RPE and used this technique to rapidly calibrate a CZ gate. The improvements in performance metrics depicted in Figures \ref{fig:angle_errors} and \ref{fig:metrics} indicate that RPE based calibration is a useful technique to fine-tune control parameters. We measured significant improvements in the coherent errors in the gate, and found that our calibration decreased the diamond distance to the target operation by almost half, which would improve the worst-case performance of the gate in running algorithmic circuits. 

Compared to alternative calibration techniques such as those based on RB metrics or process tomography metrics, RPE offers many advantages. RPE uses a small number of circuits, and for the CZ gate we require only $6(k+1)$ different circuits to achieve an estimate with bounded error of $\mathcal{O}(1/2^k)$. Achieving a similar parameter uncertainty with process tomography would require many more shots, significantly increasing the time of a calibration cycle. RB data only provides an infidelity measure that does not distinguish individual coherent errors in the gate and is only quadratically sensitive to coherent errors. In fact, our experimental results indicate that RB-based calibration would not have improved the gate performance at all because the infidelity was dominated by incoherent error from the start. 

A significant disadvantage of RPE calibration lies in the critical assumption that the effective unitary induced by the CZ pulse is well-described by an operator with support on only the $\mathbf{IZ}$, $\mathbf{ZI}$, and $\mathbf{ZZ}$ Hamiltonian terms. While our GST results indicate this is approximately true, they also indicate the presence of other smaller interactions that become important as the dominant errors on the $\mathbf{ZI}$, $\mathbf{IZ}$, and $\mathbf{ZZ}$ Hamiltonian terms are reduced. Future work will address estimating and removing these small, non-commuting contributions. 

RPE calibration offers a simple and effective technique to estimate the phases of a multi-qubit operator with Heisenberg-limited precision. The method we presented here can be directly adapted to a broad class of problems in quantum information science and engineering to enable rapid and high-precision calibration of native and compiled gates. 

\section{Code availability}
\label{sec:code}
Code demonstrating how to use the \texttt{pyRPE} \cite{pyRPE} software package for RPE analysis is available online \cite{doecode_158181}.  The experimental data presented in this paper is available upon reasonable request.

\section*{Acknowledgements} \label{sec:acknowledgements}

KCY acknowledges helpful and productive conversations with Riley Murray regarding the CMA-ES optimizer. 

This work was supported by the U.S.~Department of Energy, Office of Science, Office of Advanced Scientific Computing Research Quantum Testbed Program under Contract No.~DE-AC02-05CH11231, KCY, JPM, and KMR acknowledge support from the U.S. Department of Energy, Office of Science, Office of Advanced Scientific Computing Research Early Career Research Program. We also thank Antonio Russo for helpful technical consultations.

Sandia National Laboratories is a multi-mission laboratory managed and operated by National Technology and Engineering Solutions of Sandia, LLC (NTESS), a wholly owned subsidiary of Honeywell International Inc., for the U.S. Department of Energy’s National Nuclear Security Administration (DOE/NNSA) under contract DE-NA0003525. This written work is authored by an employee of NTESS. The employee, not NTESS, owns the right, title and interest in and to the written work and is responsible for its contents. Any subjective views or opinions that might be expressed in the written work do not necessarily represent the views of the U.S. Government. The publisher acknowledges that the U.S. Government retains a non-exclusive, paid-up, irrevocable, world-wide license to publish or reproduce the published form of this written work or allow others to do so, for U.S. Government purposes. The DOE will provide public access to results of federally sponsored research in accordance with the DOE Public Access Plan.

\appendix
\setcounter{table}{0}
\renewcommand{\thetable}{A\arabic{table}}
\setcounter{figure}{0}
\renewcommand{\thefigure}{A\arabic{figure}}

\section{Explicit formula for $n_k$.}\label{apn:n_k}
The unwinding integer $n_k$ that selects the appropriate arctangent branch in Eq.~\ref{eq:rel_phase} is, following \cite{pyRPE}, given by

\begin{equation}
n_k = \floor*{\frac{\left(\hat{\phi}_{a,b}^{(k-1)} - \tfrac{z_k}{2^k}+\tfrac{\pi}{2^k}\right)\mod 2\pi}{2\pi/2^k}},
\end{equation}
where
\begin{equation}
    z_k=\arctantwo\left( \expect{Q\ab(2^k)}_p,\expect{I\ab(2^k)}_p \right).
\end{equation}

\section{Details of the experiment design}

Table \ref{tab:circuit_table} details the explicit composition of the 6 classes of circuits used in this work and connects to the symbolic representation used in the main text. An astute reader will note that $I_{(a,b)}$ and $Q_{(a,b)}$ measurements are two-outcome measurements while full readout of an $n$-qubit register corresponds to a $2^n$-outcome measurement.  If the experimental platform only has access to full readout, what is to be done with the extra bits of information gained during measurement of the full register?  This depends on how the $I_{(a,b)}$ and $Q_{(a,b)}$ measurements are performed.  It is often---as is done in this work---easiest to emulate these measurements by following $\mathbf{U}^k$ with gates that ``unprepare'' the eigenstates of the measurement basis.  For example, for the generalized $\mathbf{X}$ measurement, $\mathbf{U}^k$ could be followed by a subcircuit that that maps (by left multiplication) $\bra{+(a,b)}$ and $\bra{-(a,b)}$ to $\bra{00}$ and $\bra{01}$.  (This is akin to how a single-qubit $\mathbf{X}$ measurement can be emulated by a $\mathbf{Z}$ measurement preceded by a Hadamard.) We would then expect the there to be no observed counts for the $\bra{10}$ and $\bra{11}$ outcomes. In such instances, indeed there \textit{isn't} any extra information to be gained by recording both bit values; all that is needed to be recorded is the value of the second recorded bit.  
\begin{table*}[]
    \centering
    \renewcommand{\arraystretch}{2}
    \begin{tabular}{|c|c|c|}
        \hline
        $\phi_{00,01}=\theta_{IZ} + \theta_{ZZ}$ &
        % circ ------------
        \begin{tikzpicture}
        \node[scale=0.75] {
        \begin{quantikz}
        \lstick{$\ket{0}$}& &  & \gate[2]{U^{2^k}_{\text{CZ}}} & & \meter{} \\
        \lstick{$\ket{0}$}& \gate{R_y(\frac{\pi}{2})} & & & \gate{R_y(-\frac{\pi}{2})} & \meter{}
        \end{quantikz} };
        \end{tikzpicture} 
        &
        % circ ------------
        \begin{tikzpicture}
        \node[scale=0.75] {
        \begin{quantikz}
        \lstick{$\ket{0}$}& &  & \gate[2]{U^{2^k}_{\text{CZ}}} & & \meter{} \\
        \lstick{$\ket{0}$}& \gate{R_y(\frac{\pi}{2})} & & & \gate{R_x(-\frac{\pi}{2})} & \meter{}
        \end{quantikz}  };
        \end{tikzpicture} 
        % circ ------------
        \\
        \hline \hline
         $\phi_{10,11} = \theta_{IZ} - \theta_{ZZ}$ 
         &
        \begin{tikzpicture}
        \node[scale=0.75] {
        \begin{quantikz}
        \lstick{$\ket{0}$}& \gate{X}  & \gate[2]{U^{2^k}_{\text{CZ}}} & \gate{X} & \meter{}\\
        \lstick{$\ket{0}$}& \gate{R_y(\frac{\pi}{2})} & &  \gate{R_y(-\frac{\pi}{2})} & \meter{}
        \end{quantikz} };
        \end{tikzpicture} 
        &
        % circ ------------
        \begin{tikzpicture}
        \node[scale=0.75] {
        \begin{quantikz}
        \lstick{$\ket{0}$}& \gate{X}  & \gate[2]{U^{2^k}_{\text{CZ}}} & \gate{X} & \meter{}\\
        \lstick{$\ket{0}$}& \gate{R_y(\frac{\pi}{2})} & &  \gate{R_x(-\frac{\pi}{2})} & \meter{}
        \end{quantikz} };
        \end{tikzpicture} 
        \\
        \hline  \hline
        $\phi_{01,11} = \theta_{ZI} - \theta_{ZZ}$
        &
        % circ ------------
        \begin{tikzpicture}
        \node[scale=0.75] {
        \begin{quantikz}
        \lstick{$\ket{0}$}& \gate{R_y(\frac{\pi}{2})} & \gate[2]{U^{2^k}_{\text{CZ}}} & \gate{R_y(-\frac{\pi}{2})} & \meter{}\\
        \lstick{$\ket{0}$}& \gate{X}  & & \gate{X} & \meter{}
        \end{quantikz} };
        \end{tikzpicture} 
        &
        % circ ------------
        \begin{tikzpicture}
        \node[scale=0.75] {
        \begin{quantikz}
        \lstick{$\ket{0}$} & \gate{R_y(\frac{\pi}{2})} & \gate[2]{U^{2^k}_{\text{CZ}}} & \gate{R_x(-\frac{\pi}{2})} & \meter{}\\
        \lstick{$\ket{0}$}& \gate{X}  & & \gate{X} & \meter{}
        \end{quantikz}};
        \end{tikzpicture} \\
        \hline
         
    \end{tabular}
    \caption{Details of the six classes of circuits used to characterize the $Z$-type phases of the controlled-Z gate. Each pair of circuits in a row implements a particular state preparation and two measurements to estimate the phase in the first column. See also Table \ref{tab:three_experiments}}
    \label{tab:circuit_table}
\end{table*}

\section{Heisenberg limited scaling}\label{apendix:Heisenberg}

Because our technique is a simple extension of single-qubit RPE in a multi-qubit setting, it is straightforward to adapt known results from the single-qubit setting in our context to prove that our multi-qubit parameter estimator is Heisenberg limited up to decoherence. We begin by arguing that each relative phase $\phi\ab$ may be estimated with Heisenberg scaling in the absence of additive errors. Next, we argue that since the model coefficients are estimated from linear combinations of the eigenphase differences, the coefficients can also be estimated with Heisenberg scaling. Finally, we comment on the effects of decoherence on the procedure. 

To argue that we can estimate the eigenphase differences $\phi\ab$ with Heisenberg scaling, we leverage Theorem 2 of \cite{Kimmel_2015} (itself adapted from \cite{Higgins2007-cc}) to assert that at depth $2^k$ the uncertainty in an estimate of the eigenphase difference $\sigma_k(\phi\ab)$ decays like $\mathcal{O}(1/2^k)$, \textit{i.e.}, is Heisenberg limited\footnote{Note that at depth $2^k$ we have actually used a gate approximately $2 \sum_{i=0}^{k} 2^i$ times, counting both the cosine and sine experiments at all the logarithmic spaced and ignoring any gate uses in state preparation and measurement. However, $2 \sum_{i=0}^{k} 2^i = 4(2^k -1)$ is still $\mathcal{O}(2^k)$}. This is by direct application of the Theorem, the only difference in our case from its application in \cite{Kimmel_2015} is that we work in a multi-qubit context, yet the dimensions of the Hilbert space are not used in the proof of the Theorem. 

Next, to argue that the uncertainty in a model coefficient is Heisenberg limited, observe that our estimator of a particular model coefficient $\theta$ is a linear combination of a set of eigenphase differences $\{\phi\ab\}$. It follows that the uncertainty in a model coefficient $\sigma_k(\theta)$ is 
\begin{equation}
    \sigma_k(\theta) = \sqrt{\sum_i \left(c_i \sigma_k(\phi_{(a_i, b_i)})\right)^2}, 
\end{equation}
where $c_i$ are constants that define the linear system. If each $\sigma_k(\phi_{(a_i, b_i)})$ decays as $\mathcal{O}(1/2^k)$, then $\sigma_k(\theta)$ decays as $\mathcal{O}(1/2^k)$. Hence, because the estimate of each $\phi_{(a_i, b_i)}$ in an experiment is Heisenberg limited, each model coefficient can also be estimated with Heisenberg scaling. 

Finally, we comment on the effect of decoherence on the scaling of our procedure. Decoherence will generally introduce errors that may be represented with an additive model. As long as these additive errors are bounded by $\sqrt{3/32}$ up to some depths $k^*$, then Theorem 1 of \cite{Kimmel_2015} may be applied to argue that each eigenphase difference $\phi_{(a, b)}$ may be estimated with Heisenberg scaling up to depth $k^*$. Per the argument of the previous paragraph, the model coefficients can also be estimated with Heisenberg scaling up to depth $k^*$. However, in the case that the additive error bounds are violated past depth $k^*$, then the Heisenberg scaling of the protocol breaks down and phase estimates after $k^*$ can no longer be said to be Heisenberg limited. The consistency checks \cite{Russo2021b} we employ will detect at which generation these bounds are violated.

\bibliography{bibliography}

\end{document}